\documentclass{article}
\usepackage{frascatiphys,here,graphicx,subfigure}
\usepackage{amssymb,amsmath}
\usepackage{graphicx}
\usepackage{subfig}
\usepackage{color}
\usepackage{slashed}
\usepackage{enumerate}
\begin{document}
\title{ 
Minimal seesaw models and minimal lepton flavor violation
}
\author{
Audrey Deg\'{e}e \\
{\em IFPA, Dep. AGO,
      Universit\'{e} de Li\`{e}ge,}\\
{\em Bat B5, Sart Tilman B-4000
      Liege 1, Belgium}
}
\maketitle
\baselineskip=11.6pt
\begin{abstract}
 We study the implications of the global $U(1)_R$ symmetry present in
  minimal lepton flavor violating implementations of the seesaw mechanism for neutrino masses. Our discussion
is done in the context of explicit minimal type-I seesaw scenarios, where depending on the R-charge
assignment different models can be constructed. We study the charged lepton flavor violating
phenomenology of a concrete realization paying special attention to
$\mu\to e\gamma$ and $\mu\to 3e$.

\end{abstract}
\baselineskip=14pt
\section{Introduction}
\label{sec:intro}
The observation of neutrino flavor oscillations constitutes an
experimental proof of lepton flavor violation\cite{Schwetz:2011zk}. 
In principle, other manifestations of such
effects could be expected to show up in the charged lepton sector as
well. However, the lack of a definitive model for neutrino mass
generation implies that conclusive predictions for lepton flavor
violating processes can not be made, and even assuming a concrete
model realization for neutrino masses, predictions for such effects
can only be done if the flavor structure of the corresponding
realization is specified.

In this regards the minimal flavor violating hypothesis
\cite{Chivukula:1987py,Hall:1990ac,D'Ambrosio:2002ex} is a very useful
guide for constructing predictive models in which lepton violating signals
are entirely determined by the low-energy neutrino data.
However minimal lepton flavor violation (MLFV) can not be
uniquely implemented and depends on the new physics responsible for
neutrino masses. 
Here considering a type-I seesaw mechanism i.e. taking the new degrees of freedom to be heavy fermionic electroweak singlets
(right-handed (RH) neutrinos), we study the implications of the $U(1)_R$ present in MLFV 
models assuming it is slightly broken.
The full analysis is done in the context of a minimal type-I seesaw setup (2 RH neutrinos) where
the number of parameters and low energy observables are such that all flavor
effects are entirely determined by neutrino observables up to normalizations factors.

\section{The setups}
\label{sec:review}
The kinetic Lagrangian of the standard model extended with two RH
neutrinos exhibits the global $G=U(3)_e\times U(3)_\ell\times U(2)_N$
symmetry. This group can be rewritten as $U(1)_Y\times U(1)_L\times U(1)_R\times
G_F$ where $U(1)_{Y,L}$ can be identified with global hypercharge and
lepton number whereas the $U(1)_R$ is a ``new'' global
symmetry, already mentioned in the introduction\cite{D'Ambrosio:2002ex,Alonso:2011jd}.

The charges associated with this global transformation (hereafter denoted by $R$)
are arbitrary, and thus different $R$-charge assignments define
different models. 
Here we will explicitly consider the seesaw Lagrangian with a slightly broken
$U(1)_R$ and discuss a generic model\footnote{A second class of models can be constructed in which the small breaking of
$U(1)_R$ allows to decouple the lepton number breaking scale from the
RH neutrino mass scale. But this decoupling implies also
a suppression of the corresponding Yukawa couplings, thus leading to non-observable
charged lepton flavor violating effects. For more details see\cite{AristizabalSierra:2012yy}} that
arises from a particular R-charge assignment.

Precisely speaking we refer to the following:
$R(N_1,\ell_i,e_i)=+1$, $R(N_2)=-1$, $R(H)=0$ where $\ell_i,e_i$ and $N_{1,2}$ are respectively the electroweak lepton doublets, singlets and RH neutrinos.
The Lagrangian is thus given by
\begin{equation}
  \label{eq:seesaw-lag-N!N2mismatch}
  {\cal L}=  
  - \bar \ell\,\pmb{\lambda_1}^*\,N_1 \tilde H
  - \epsilon_\lambda\, \bar \ell\,\pmb{\lambda_2}^*\,N_2 \tilde H
  - \frac{1}{2} N_1^T\,C M\,N_2
  - \frac{1}{2}\epsilon_N N_a^T\,C M_{aa}\,N_a + 
  \mbox{h.c.}\,.
\end{equation}
The dimensionless $\epsilon_{\lambda,N}$ parameters determine the amount of $U(1)_R$
breaking and thus are tiny, $\tilde H = i\sigma_2 H^*$, $C$ is the charge conjugation
operator, the $\pmb{\lambda_a}$'s are Yukawa vectors in flavor space and $M$ and $M_{aa}$
are the parameters that define the RH neutrino mass matrix. The diagonalization of the Majorana RH
neutrino mass matrix leads to two quasi-degenerate states with masses
given by $  M_{N_{1,2}}=M\mp\frac{M_{11}+M_{22}}{2}\epsilon_N $ and 
in the basis in which the RH neutrino mass matrix is diagonal the
Yukawa couplings $\lambda_{ka}$ become $-\frac{(i)^a}{\sqrt{2}}\left[\lambda_{k1}
    + (-1)^a\epsilon_\lambda\lambda_{k2}\right]$ 
$(k=e,\mu,\tau\;\;\mbox{and}\;\; a=1,2)\,.$
In terms of these redefined Yukawa couplings, the effective light neutrino mass matrix, up to ${\cal
  O}(\epsilon_N\epsilon_\lambda^2)$, is given by
\begin{equation}
  \label{eq:light-nmm-N1N2mismatch}
  \pmb{m_\nu^\text{eff}}=-\frac{v^2\epsilon_\lambda}{M}
  |\pmb{\lambda_1}||\pmb{\Lambda}|
  \left(
    \pmb{\hat \lambda_1}^*\otimes\pmb{\hat \Lambda}^* 
    +
    \pmb{\hat \Lambda}^*\otimes\pmb{\hat \lambda_1}^*
  \right),\end{equation}
with $\pmb{\hat \Lambda}^*=\pmb{\hat \lambda_2}^*-
  \frac{M_{11}+M_{22}}{4M}\frac{\epsilon_\lambda}{\epsilon_N}
  \pmb{\hat \lambda_1}^*\,.$ Note that we have expressed the parameters space vector
$\pmb{\lambda_1}$ and $\pmb{\Lambda}$ in terms of their unitary vector $\pmb{\hat \lambda_1}$, $\pmb{\hat \Lambda}$ and moduli $|\pmb{\lambda_1}|$, $|\pmb{\Lambda}|$.
Since $\epsilon_\lambda\ll 1$ small neutrino masses do not require
heavy RH neutrinos or small Yukawa couplings, thus potentially
implying large lepton flavor violating effects. 

It turns out that due to the structure of
\eqref{eq:light-nmm-N1N2mismatch} the vectors $\pmb{\lambda_1}$
and $\pmb{\Lambda}$ can be entirely determined by means of the solar
and atmospheric mass scales and mixing angles, up to the factors
$|\pmb{\lambda_1}|$ and $|\pmb{\Lambda}|$. The expression depend on the light
neutrino mass spectrum, in the normal case they read\cite{Gavela:2009cd}
\begin{align}
    \label{eq:Yukawas-normalS1}
    \pmb{\lambda_1}&=|\pmb{\lambda_1}|\;\pmb{\hat \lambda_1}=
    \frac{|\pmb{\lambda_1}|}{\sqrt{2}}
    \left(
      \sqrt{1+\rho}\,\pmb{U_3}^* + \sqrt{1-\rho}\,\pmb{U_2}^*
    \right)\,,\\
    \label{eq:Yukawas-normalS2}
    \pmb{\Lambda}&=|\pmb{\Lambda}|\;\pmb{\hat \Lambda}=
    \frac{|\pmb{\Lambda}|}{\sqrt{2}}
    \left(
      \sqrt{1+\rho}\,\pmb{U_3}^* - \sqrt{1-\rho}\,\pmb{U_2}^*
    \right)\,,
  \end{align}
where $\pmb{U_i}$ denote the columns of the leptonic mixing matrix and
\begin{equation}
  \label{eq:rho-and-r-NS}
  \rho=\frac{\sqrt{1+r}-\sqrt{r}}{\sqrt{1+r}+\sqrt{r}}\,,\qquad
  r=\frac{m_{\nu_2}^2}{m_{\nu_3}^2-m_{\nu_2}^2}\,.
\end{equation}

\section{Lepton flavor violating  processes}
\label{sec:lfv-processes}
With potentially large Yukawa couplings and RH neutrinos in the TeV ballpark, 
charged lepton flavor violating processes could be expected to have large rates.
In what follows we analyze $l_i\to l_j\gamma$ and $l_i^-\to l_j^- l_j^- l_j^+$ .
\subsection{$l_i\to l_j\gamma$  processes}
The decay branching ratios can be approximated by\cite{AristizabalSierra:2012yy}:

\begin{equation}
  \label{eq:4}
  \text{BR}(l_i \to l_j \gamma)\simeq\frac{\alpha}{1024 \pi^4}
  \frac{m_i^5}{M^4}\frac{|\pmb{\lambda_1}|^4}{\Gamma_\text{Tot}^{l_i}}
  \left|
    \hat\lambda_{i1}\;\hat\lambda_{j1}^*
  \right|^2\,,
\end{equation}
where $\Gamma_\text{Tot}^{l_i}$ stands for the total decay width of the corresponding decaying 
charged lepton $l_i$.

Among these lepton flavor violating processes, presently
the $\mu\to e\gamma$ transition is the most severely constrained. The MEG collaboration recently established
an upper bound of $2.4\times 10^{-12}$ at the 90\%
C.L.~\cite{Adam:2011ch}. So from now on we focus on that process.
To quantify this effect, we randomly generate low energy observables in their $2\sigma$ ranges and the
parameters $|\pmb{\lambda_1}|$ and $M$ in the ranges
$[10^{-5},1]$ and $[10^2,10^6]$ GeV allowing RH neutrino
mass splittings in the range $[10^{-8},10^{-6}]$ GeV. µ

\begin{figure}[t!]
    \begin{center}
        {\includegraphics[scale=0.95]{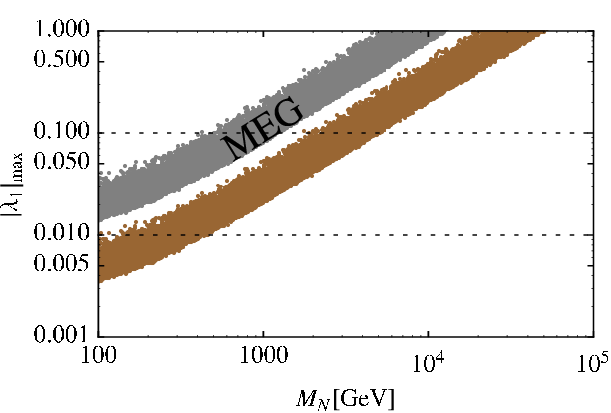}}
        {\includegraphics[scale=0.95]{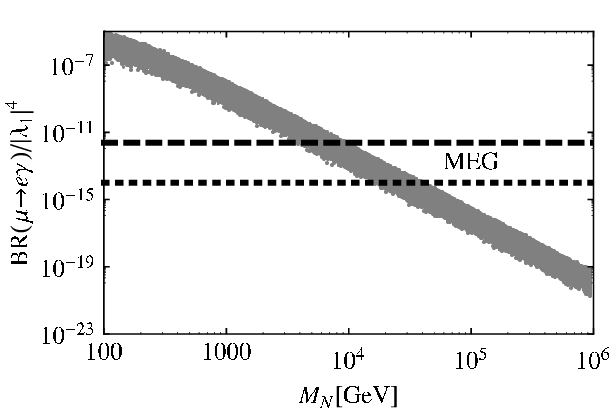}}
        \caption{\it Radiative LFV decay branching ratio BR$(\mu\to
    e\gamma)$ for normal light neutrino mass spectra as a function of the common RH
    neutrino mass. The upper horizontal dashed line indicates the
    current limit on BR$(\mu\to e\gamma)$ from the MEG experiment
    \cite{Adam:2011ch}, whereas the lower one the future experimental
    sensitivities\cite{meg-futureS}.}
  \label{fig:radiative-lfv}
    \end{center}
\end{figure}
The results for the normal mass spectrum case are
displayed in figure \ref{fig:radiative-lfv}.
It can be seen that BR$(\mu\to e\gamma)$ can
reach the current experimental limit reported by the MEG experiment
\cite{Adam:2011ch} for RH neutrino masses $M_N<0.1~\mathrm{TeV},1~\mathrm{TeV},~10$~TeV provided $|{\pmb{\lambda_1}}|\gtrsim 2\times 10^{-2},~10^{-1}, ~1$, respectively.

\subsection{$l_i^-\to l_j^- l_j^- l_j^+$ processes}
\begin{figure}[t!]
    \begin{center}
        {\includegraphics[scale=0.95]{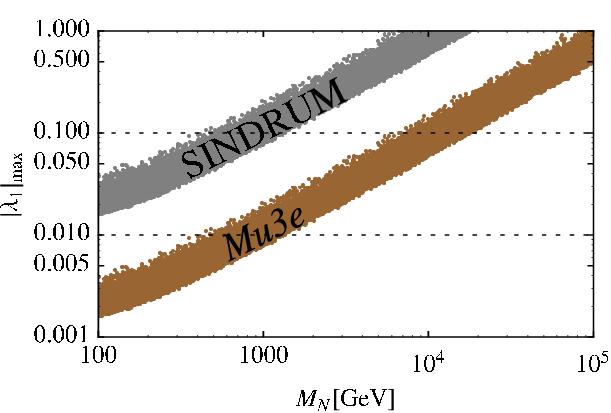}}
        {\includegraphics[scale=0.95]{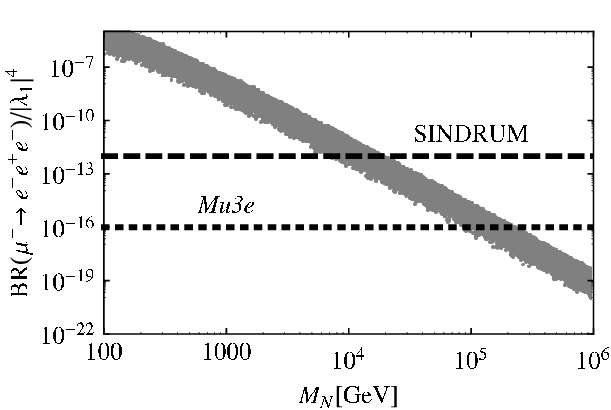}}
        \caption{\it Decay branching ratio BR($\mu^-\to e^- e^+ e^-$)
    for normal light neutrino mass spectra as a function of common RH neutrino mass. The
    upper horizontal dashed line indicates the current bound for
    $\mu^-\to e^+e^-e^-$ placed by the SINDRUM experiment
    \cite{Bellgardt:1987du}, whereas the lower one future experimental
    sensitivities \cite{psi}.}
  \label{fig:muto3elec}
    \end{center}
\end{figure}

The decay branching ratios for these processes have been
calculated in~\cite{Ilakovac:1994kj,AristizabalSierra:2012yy}. The most constrained process in
this case is $\mu^-\to e^+e^-e^-$ for which the SINDRUM experiment has
placed a bound on the decay branching ratio of $10^{-12}$ at 90\%
C.L.~\cite{Bellgardt:1987du}.
Following the same numerical procedure used in the previous section we calculate the
decay branching ratio for $\mu^-\to e^+e^-e^-$. The results are shown in figure \ref{fig:muto3elec}.
Again, as in the $\mu \to e \gamma$ case, it can be seen that the decay branching ratio can 
saturate the current experimental bound for RH neutrino masses 
$M_N<0.1~\mathrm{TeV},1~\mathrm{TeV},~10$~TeV provided $|{\pmb{\lambda_1}}|\gtrsim 2\times 10^{-2},~10^{-1}, ~1$, 
respectively.
\section{Conclusion}
The presence of an extra global $U(1)_R$ in the seesaw Lagrangian allows the construction
of different types of models, all of them determined by the R-charge assignments of the lepton
sector. We have considered a concrete realization and analyzed its consequences for the most promising 
charged lepton flavor violating decays, namely $\mu \to e \gamma$ and $\mu^-\to e^- e^+ e^-$.
Our analysis shows that for a large mass range of the lepton flavor violating mediators these processes might be observed in near future experiments.

\section{Acknowledgements}
I would like to thank D. Aristizabal Sierra and J. F. Kamenik for the fruitful collaboration. D. Aristizabal Sierra also
for the help during the preparation of this work.

\end{document}